\documentclass[doublecol]{epl2}
\usepackage{mathtools}

\title{Ornstein-Uhlenbeck information swimmers with external and internal feedback controls}
\shorttitle{Ornstein-Uhlenbeck information swimmers} 

\author{Zhanglin Hou\inst{1,2} \and Ziluo Zhang\inst{1,3} \and Jun Li\inst{4,1} \and Kento Yasuda\inst{5} \and Shigeyuki Komura\inst{1,2}}
\shortauthor{Z. Hou \etal}

\institute{                    
  \inst{1} Wenzhou Institute, University of Chinese Academy of Sciences, Wenzhou, Zhejiang 325001, China\\
  \inst{2} Oujiang Laboratory, Wenzhou, Zhejiang 325000, China \\
  \inst{3} Institute of Theoretical Physics, Chinese Academy of Sciences, Beijing 100190, China \\
  \inst{4} Department of Physics, Wenzhou University, Wenzhou, Zhejiang 325035, China \\
  \inst{5} Research Institute for Mathematical Sciences, Kyoto University, Kyoto 606-8502, Japan \\
}

\abstract{
Using an underdamped active Ornstein-Uhlenbeck particle, we propose two information swimmer models having 
either external or internal feedback control and perform their numerical simulations. 
Depending on the velocity that is measured after every fixed time interval (measurement time), the friction coefficient is 
modified in the externally controlled model, whereas the persistence time for the activity is changed in the internally 
controlled one. 
In the steady state, both of these information swimmers acquire finite average velocities in the noisy environment, 
and their efficiencies can be maximized by tuning the measurement time.
The internally controlled swimmer can generally achieve a larger velocity and efficiency than the externally 
controlled one when the active fluctuation is large.
}

\begin{document}

\maketitle

\section{Introduction}

In the studies of active matter, much interest has been focused on the collective motions of 
self-propelled particles~\cite{Marchetti13}. 
It is also important to understand the mechanism of single-particle motion from 
the perspective of non-equilibrium statistical mechanics~\cite{Hosaka22}.
Among various models, the self-propulsion of an active Brownian particle (ABP) is generated by an 
internal driving force combined with overdamped orientational Brownian dynamics~\cite{Romanczuk12}.
An overdamped self-propelled motion can be described by an active Ornstein-Uhlenbeck particle (AOUP) 
driven by a stochastic force whose memory decays exponentially in time~\cite{Uhlenbeck30}.
Since the AOUP model exhibits persistent particle motion mimicking the activity, it offers a basic reference 
for active dynamics of cells and bacteria~\cite{Martin21}.

Although both ABP and AOUP are active, they undergo Brownian dynamics in the long time 
limit~\cite{tenHagen11} and cannot have any net locomotion on average.
For microswimmers in a Newtonian fluid, it is known that non-reciprocal cyclic body motion is required for their 
persistent locomotion~\cite{Najafi04,Yasuda23}.
To maintain such continuous movement, a constant supply of mechanical energy and its dissipation 
to the surrounding fluid is necessary~\cite{Golestanian08}. 
Other active systems driven by energy injection include Janus particles that consume a chemical 
fuel~\cite{Howse07}.  
In contrast to these microswimmers, the ``odd microswimmer" consisting of three spheres and two odd 
springs is purely driven by thermal energy of the surrounding fluid~\cite{Yasuda21,Li24}.

\begin{figure*}[tbh]
\centering
\includegraphics[scale=0.23]{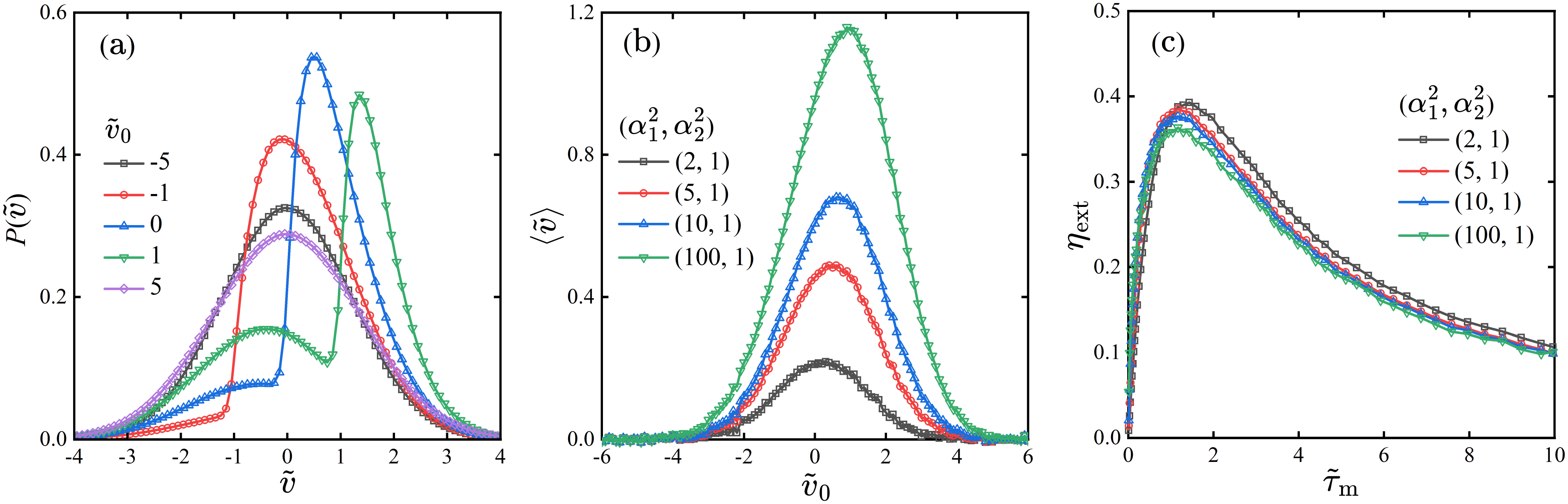}
\caption{
Various statistical properties of the E-OUIS model with external feedback control [see Eqs.~(\ref{externalmodel})--(\ref{externalmodelnoisexi})].
The fixed parameters are $\tilde \tau_{\rm a} = 1$ and $\tilde A = 1$.
(a) The steady-state velocity distribution function $P(\tilde v)$ when the threshold velocity is changed within the range of $-5 \le \tilde{v}_0 \le 5$.
Here, we choose $(\alpha^2_1,\alpha^2_2)=(10,1)$ and the measurement time interval is $\tilde \tau_{\rm m} = 0.01$.
(b) The steady-state average velocity $\langle \tilde v \rangle$ as a function of the threshold velocity $\tilde v_0$ when 
$(\alpha^2_1,\alpha^2_2)=(2,1)$, $(5,1)$, $(10,1)$, $(100,1)$ and $\tilde \tau_{\rm m} = 0.01$.
(c) The efficiency $\eta_{\rm ext}$ of the E-OUIS model [see Eq.~(\ref{efficiencyII})] as a function of 
$\tilde \tau_{\rm m}$ for the same combinations of $(\alpha^2_1,\alpha^2_2)$ shown in (b). 
The threshold velocity is $\tilde v_0 = 0$.
}
\label{FIG:thermal_OUIS}
\end{figure*}

Recently, much attention has been paid to active systems that utilize information instead of energy, i.e., 
informational active matter~\cite{VanSaders24}.
During the last decade, there have been many studies on ``information engines" that use information to extract 
mechanical work~\cite{Toyabe10,Paneru20,Paneru22}.
One example proposed by Huang \textit{et al.} is the ``information swimmer" 
in which the swimmer periodically measures its velocity and adjusts its friction coefficient~\cite{Huang20}. 
They showed that the information swimmer can achieve a steady-state velocity without external energy input, 
which is consistent with the extended second law of thermodynamics with information~\cite{Parrondo15}.
This model can also be regarded as one type of ``information ratchet" that leads to 
directional transport or locomotion by repeating measurements and feedback controls~\cite{Serreli07,Hwang19}.

In the absence of measurement and feedback, the information swimmer proposed by 
Huang \textit{et al.}~\cite{Huang20} is passive and purely governed by thermal fluctuations. 
From the viewpoint of informational active matter~\cite{VanSaders24}, however, it is of interest to consider 
a swimmer that controls its internal activity depending on the measurement result.
In this Letter, we propose models of information swimmers by using an underdamped AOUP and call them 
``Ornstein-Uhlenbeck information swimmers" (OUISs). 
To highlight the role of activity, we discuss two different models: (i) OUIS with \textit{external} 
feedback control (E-OUIS) and (ii) OUIS with \textit{internal} feedback control (I-OUIS). 
Depending on the particle velocity, the friction coefficient is adjusted in the E-OUIS (similar to Ref.~\cite{Huang20}), 
whereas the persistence time for the activity is controlled in the I-OUIS.  
While the externally controlled swimmer requires some structural changes in volume or shape~\cite{Huang20}, 
the internal control can be achieved by various chemical and mechanical processes~\cite{Tkacik16}, 
such as the run-and-tumble motion of \textit{Escherichia coli}~\cite{Kurzthaler24,Zhao24}.  
We perform numerical simulations of the two models and discuss their swimming performance.
Although both OUIS models can maintain a steady motion, the I-OUIS can generally achieve a larger velocity 
and efficiency than the E-OUIS when the active fluctuation becomes large.

\section{OUIS with external feedback control (E-OUIS)}

We consider an underdamped AOUP moving in a one-dimensional space~\cite{Caprini21,Nguyen22}.
In the E-OUIS model, the AOUP measures its center-of-mass velocity $v$ after every time 
interval $\tau_{\rm m}$ (measurement time) and compares it with a threshold velocity $v_0$~\cite{Huang20,Garcia19}. 
The friction coefficient is set to be $\alpha_1^2 \Gamma$ and $\alpha_2^2 \Gamma$ if $v \leq v_0$ and $v > v_0$, 
respectively, where $\alpha_1$ and $\alpha_2$ are dimensionless coefficients and we typically choose 
$\alpha_1 > \alpha_2$.
The change in the friction coefficient can be induced by some structural changes, such as the particle volume, 
shape, and surface structure~\cite{Huang20}.
The results of measurement are recorded in the swimmer's internal memory that is assumed to be sufficiently large.

During the time interval $n \tau_{\rm m} < t < (n+1) \tau_{\rm m}$, the Langevin equation for the underdamped 
E-OUIS is given by 
\begin{align}
& M\dot{v}(t) 
\nonumber \\ 
& =  \begin{dcases}
		-\alpha_1^2 \Gamma [v(t)-u(t)]+\alpha_1 \sqrt{2 \Gamma k_{\rm B}T} \zeta (t),  \quad v(n \tau_{\rm m}) \leq v_0, \\
		-\alpha_2^2 \Gamma [v(t)-u(t)]+\alpha_2 \sqrt{2 \Gamma k_{\rm B}T} \zeta (t),  \quad v(n \tau_{\rm m}) > v_0.
	\end{dcases}
\label{EQ:model2}
\end{align}
Here, $M$ is the particle mass, $\dot{v}=dv/dt$, $\Gamma$ is the friction coefficient, 
$k_{\rm B}$ is the Boltzmann constant, $T$ is the temperature, and $\zeta(t)$ is the Gaussian white noise 
with zero mean and unit variance:
\begin{equation}
\langle \zeta (t) \rangle =0, \quad \langle \zeta (t) \zeta (t^\prime) \rangle=\delta (t-t^\prime).
\label{EQ:white noise}
\end{equation}
The coefficients of the noise terms in Eq.~(\ref{EQ:model2}) are chosen to satisfy the fluctuation-dissipation 
relation for both $v \leq v_0$ and $v > v_0$, reflecting the equilibrium nature of the ambient 
environment~\cite{DoiBook}. 
Note that the threshold velocity $v_0$ can take both positive and negative values. 
In the above, we assume that the change of the friction coefficient does not require any mechanical work~\cite{Huang20}.

In Eq.~(\ref{EQ:model2}), $u$ is a stochastic driving velocity with memory on a finite time leading to a persistent 
motion~\cite{Caprini21,Nguyen22}.
It mimics the activity of the particle and evolves through an Ornstein-Uhlenbeck process~\cite{Uhlenbeck30}
\begin{equation}
\dot{u}(t)=-\frac{u(t)}{\tau_{\rm a}}+ \frac{\sqrt{2A}}{\tau_{\rm a}} \xi(t),
\label{EQ:coloured noise}
\end{equation}
where $\tau_{\rm a}$ is the persistence time, $A$ is the strength of active fluctuation (having the dimension 
of a diffusion constant).
The term $\xi(t)$ in Eq.~(\ref{EQ:coloured noise}) also represents the Gaussian white noise with zero mean and 
unit variance:
\begin{equation}
\langle \xi (t) \rangle =0, \quad \langle \xi (t) \xi (t^\prime) \rangle=\delta (t-t^\prime).
\label{EQ:white noise2}
\end{equation}
Although we do not discuss here the chemical or mechanical origins of the driving velocity $u$, 
it pushes the particle away from equilibrium and guarantees a persistent motion in one direction for times smaller 
than $\tau_{\rm a}$.
The parameters $\tau_{\rm a}$ and $A$ can be used to quantify the activity of the AOUP, such as the persistence 
length $\ell_0=\sqrt{A \tau_{\rm a}}$ and the magnitude of the driving velocity 
$u_0=\sqrt{A/ \tau_{\rm a}}$~\cite{Caprini21,Nguyen22}.

In the following analysis, we use the Brownian relaxation time $\tau=M/\Gamma$ and 
the thermal velocity $v_T=\sqrt{k_{\rm B}T/M}$ 
to rescale the variables as 
$\tilde t = t/\tau$, $\tilde v(\tilde t) = v(t)/v_T$, 
$\tilde u(\tilde t) = u(t)/v_T$,
$\tilde \zeta (\tilde t) = \sqrt{2M/\Gamma} \zeta(t)$, and 
$\tilde \xi (\tilde t) = \sqrt{2M/\Gamma} \xi(t)$. 
Moreover, the dimensionless parameters are defined as 
$\tilde \tau_{\rm m}=\tau_{\rm m}/\tau$, 
$\tilde v_0 = v_0/v_T$, 
$\tilde \tau_{\rm a}=\tau_{\rm a}/\tau$, and
$\tilde A = A \Gamma /(k_{\rm B}T)$.
Then, the dimensionless form of the E-OUIS model can be summarized as 
\begin{gather}
\frac{d\tilde v(\tilde t)}{d\tilde t} = \begin{dcases}
-\alpha_1^2 [\tilde v(\tilde t) - \tilde u(\tilde t)]+\alpha_1 \tilde \zeta(\tilde t), & \quad \tilde v(n \tilde \tau_{\rm m}) \leq \tilde v_0, 
\\
-\alpha_2^2 [\tilde v(\tilde t) - \tilde u(\tilde t)]+\alpha_2 \tilde \zeta(\tilde t), & \quad \tilde v(n \tilde \tau_{\rm m}) > \tilde v_0,
\end{dcases}
\label{externalmodel}
\\
\frac{d \tilde u (\tilde t)}{d \tilde t}=-\frac{\tilde u(\tilde t)}{\tilde \tau_{\rm a}} + \frac{\sqrt{\tilde A}}{\tilde \tau_{\rm a}} \tilde \xi(\tilde t),
\label{EQ:colourednondimension}
\\
\langle \tilde \zeta(\tilde t) \rangle=0, \quad \langle \tilde \zeta(\tilde t) \tilde \zeta(\tilde t^\prime) \rangle
=2 \delta(\tilde t - \tilde t^\prime),
\\
\langle \tilde \xi(\tilde t) \rangle=0, \quad \langle \tilde \xi(\tilde t) \tilde \xi(\tilde t^\prime) \rangle
=2 \delta(\tilde t - \tilde t^\prime),
\label{externalmodelnoisexi}
\end{gather}
for $n \tilde \tau_{\rm m} < \tilde t < (n+1) \tilde \tau_{\rm m}$.

When $\alpha_1=\alpha_2=1$, namely, when the feedback control is absent, 
one can easily convert the above Langevin equations to a Fokker-Planck equation and obtain the velocity 
autocorrelation function of an AOUP, as explained in the Supplementary Information (SI).
Using the equal-time velocity correlation, one can define the effective temperature of an AOUP as 
$k_{\rm B}T^\ast = M \langle v^2 \rangle$, where
\begin{equation}
k_{\rm B} T^\ast=k_{\rm B} T + \frac{A\Gamma \tau}{\tau+ \tau_{\rm a}}
=k_{\rm B} T \left( 1+\frac{\tilde A}{1+\tilde \tau_{\rm a}} \right).
\label{effT}
\end{equation}
Hence, the effective temperature $T^\ast$ is larger than $T$, and the additional term is proportional to the 
active fluctuation strength $A$. 
In the absence of feedback control ($\alpha_1=\alpha_2$), however, the E-OUIS does not exhibit any net locomotion 
in the long time limit, as mentioned before.

The above stochastic equations for the E-OUIS are discretized over the small time step $\Delta \tilde t = 0.001$, 
and they are numerically integrated by using the first-order Euler-Maruyama scheme~\cite{Kloeden92}.
Among several model parameters, we fix $\tilde \tau_{\rm a}=1$ and $\tilde A=1$, while we change 
$\alpha_1^2$, $\alpha_2^2$, $\tilde \tau_{\rm m}$, and $\tilde v_0$. 
When the particle has reached the steady state after $10^7$ time steps, we obtain the time-independent 
velocity distribution function $P(\tilde v)$ to calculate the steady-state average velocity 
$\langle \tilde v \rangle = \int_{-\infty}^{\infty} d\tilde v \, \tilde v P(\tilde v)$.
On the other hand, the steady-state distribution function $P(\tilde u)$ of the driving velocity 
$u$ is purely Gaussian [see Eqs.~(\ref{EQ:colourednondimension}) and (\ref{externalmodelnoisexi})], and its 
average should vanish, $\langle \tilde u \rangle=0$, as we have checked numerically.

In Fig.~\ref{FIG:thermal_OUIS}(a), we show the results of the velocity distribution function 
$P(\tilde v)$ of the E-OUIS when $(\alpha_1^2, \alpha_2^2, \tilde \tau_{\rm m}) = (10, 1, 0.01)$ (corresponding to 
$\tilde \tau_{\rm m}=10 \Delta \tilde t$) and $\tilde v_0$ is varied between $-5 \le \tilde v_0 \le 5$.
When the absolute value of $\tilde v_0$ is as large as $\vert \tilde v_0 \vert \approx 5$, $P(\tilde v)$ is almost 
Gaussian and $\langle \tilde v \rangle$ vanishes. 
For $\vert \tilde v_0 \vert < 4$, however, $P(\tilde v)$ becomes highly asymmetric and even bimodal 
for $\tilde v_0>0$ (see later discussion for its physical origin). 
In such a situation, the average velocity is finite, $\langle \tilde v \rangle >0$, and hence the E-OUIS 
model acquires net locomotion under noisy environment.  
In Fig.~\ref{FIG:thermal_OUIS}(b), we plot $\langle \tilde v \rangle$ as a function of $\tilde v_0$ for 
different values of $\alpha_1^2$. 
The maximum of $\langle \tilde v \rangle$ increases for larger $\alpha_1^2$, and it occurs at 
positive $\tilde v_0$ such as at $\tilde v_0 \approx 0.6$ when $\alpha_1^2=10$.
Notice that the induced average velocity is the order of thermal velocity $v_T$.

Next, we argue the swimming performance of the E-OUIS model. 
First, the change rate of information entropy is given by $\dot{I}=I / \tau_{\rm m}$, 
where $I$ is the mutual information~\cite{Parrondo15}. 
Since there is no error in the measurement, the mutual information is equal to the Shannon entropy,
$I = - k_{\rm B} \sum_i p_i \ln p_i$, where $p_i$ is the probability of the state $i=1,2$ and satisfies $\sum_i p_i=1$. 
We have assumed that the swimmer's memory space is sufficiently large so that the information entropy can increase steadily, 
and hence there is no need to erase information~\cite{Huang20}. 
Second, we estimate the power of the swimmer by the product of the frictional force 
$\alpha_i^2 \Gamma \langle v-u \rangle_i$ and the center-of-mass velocity $\langle v \rangle_i$, and then  
averaged over the two states $i=1,2$.
Dividing the average power by the effective temperature $T^\ast$ in Eq.~(\ref{effT}),
we estimate the entropy production rate due to the frictional motion by  
$\dot{\sigma}_v=\sum_i \alpha_i^2 \Gamma \langle v-u \rangle_i \langle v \rangle_i p_i/T^\ast$.
Then, we define the entropic efficiency of the E-OUIS model as 
\begin{align}
\eta_{\rm ext} = \frac{\dot{\sigma}_v}{\dot{I}} 
=\frac{\sum_i \alpha_i^2 \langle \tilde v -\tilde u \rangle_i \langle \tilde v \rangle_i p_i/\tilde{T}^\ast}{\tilde{I}/ \tilde \tau_{\rm m}},
\label{efficiencyII}
\end{align}
where $\tilde{T}^\ast=T^\ast/T$ and $\tilde{I}=I/k_{\rm B}$.
Huang \textit{et al.}\ considered a similar efficiency for an information swimmer that can maintain directional motion 
only by measurement and feedback~\cite{Huang20}.

The above efficiency $\eta_{\rm ext}$ is plotted as a function of $\tilde \tau_{\rm m}$ in 
Fig.~\ref{FIG:thermal_OUIS}(c) when $\tilde v_0 = 0$. 
The overall behavior is similar for different choices of $\alpha_1^2$.
The efficiency is low for small and large $\tilde \tau_{\rm m}$ values, taking a maximum value 
$\eta_{\rm ext} \approx 0.38$ at around $\tilde \tau_{\rm m} \approx 1$ for all $\alpha_1^2$.
In other words, the efficiency is maximized when the measurement time is close to the Brownian 
relaxation time $\tau$. 
As separately shown in Fig.~S1 of the SI, the velocity distribution  
$P(\tilde v)$ for $\tilde \tau_{\rm m} =1$ is more symmetric and $\langle \tilde v \rangle$ is smaller than 
that for $\tilde \tau_{\rm m} =0.01$.
However, since $\dot{I}$ is also smaller for $\tau_{\rm m}=1$, $\eta_{\rm ext}$ is also maximized at 
around $\tilde \tau_{\rm m} \approx 1$.
We have further checked the other cases of $(\alpha_2^2, \tilde \tau_{\rm m}) = (0.1, 0.01)$ and systematically changed 
$\alpha_1^2$, as shown in Fig.~S2.
The results are similar to those in Fig.~\ref{FIG:thermal_OUIS}, showing a robust dependence on the model parameters.
However, as shown in Fig.~S2(c), $\eta_{\rm ext}$ decreases much slower when $\tau_{\rm m}$ is made larger.

\begin{figure*}[tbh]
\centering
\includegraphics[scale=0.23]{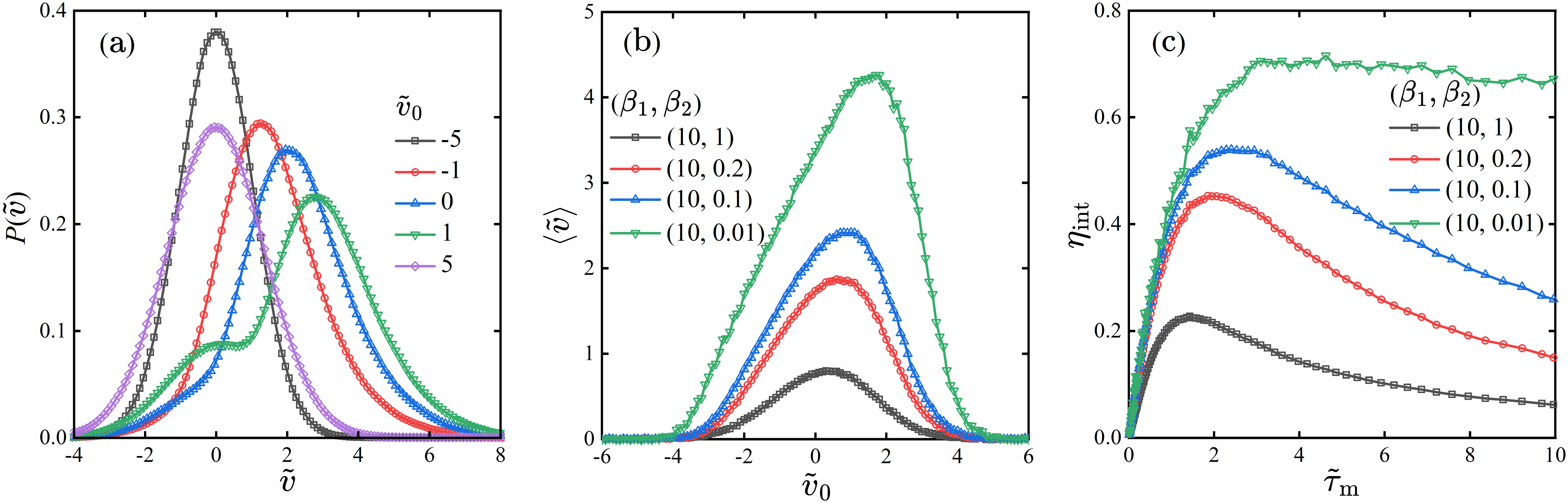}
\caption{
Various statistical properties of the I-OUIS model with internal feedback control [see Eqs.~(\ref{internalmodelnondim})--(\ref{internalmodelxi})].
The fixed parameters are $\tilde \tau_{\rm a} = 1$ and $\tilde A = 1$.
(a) The steady-state velocity distribution function $P(\tilde v)$ when the threshold velocity is changed within the range of $-5 \le \tilde{v}_0 \le 5$.
Here, we choose $(\beta_1,\beta_2)=(10,0.1)$ and the measurement time interval is $\tilde \tau_{\rm m} = 0.01$.
(b) The steady-state average velocity $\langle \tilde v \rangle$ as a function of the threshold velocity $\tilde v_0$ when 
$(\beta_1,\beta_2)=(10,1)$, $(10,0.2)$, $(10,0.1)$, $(10,0.01)$ and $\tilde \tau_{\rm m} = 0.01$.
(c) The efficiency $\eta_{\rm int}$ of the I-OUIS model [see Eq.~(\ref{EQ:efficiencyIII_modified})] as a function of 
$\tilde \tau_{\rm m}$ for the same combinations of $(\beta_1,\beta_2)$ shown in (b). 
The threshold velocity is $\tilde v_0 = 0$.
}
\label{FIG:non-thermal OUIS}
\end{figure*}

\begin{figure*}[tbh]
\centering
\includegraphics[scale=0.23]{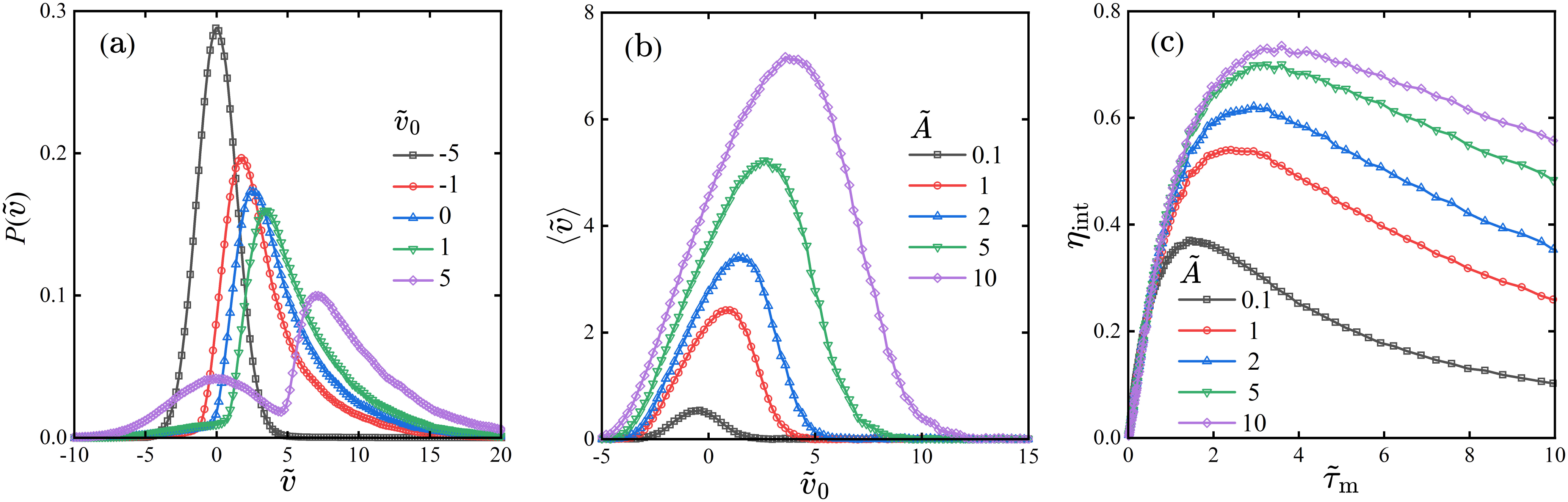}
\caption{
Various statistical properties of the I-OUIS model with internal feedback control [see Eqs.~(\ref{internalmodelnondim})--(\ref{internalmodelxi})].
The fixed parameters are $\tilde \tau_{\rm a} = 1$ and $(\beta_1,\beta_2)=(10, 0.1)$.
(a) The steady-state velocity distribution function $P(\tilde v)$ when the threshold velocity is changed within the range of $-5 \le \tilde{v}_0 \le 5$.
Here, we choose $\tilde A = 10$ and the measurement time interval is $\tilde \tau_{\rm m} = 0.01$.
(b) The steady-state average velocity $\langle \tilde v \rangle$ as a function of the threshold velocity $\tilde v_0$ when 
$\tilde A =0.1$, $1$, $2$, $5$, $10$ and $\tilde \tau_{\rm m} = 0.01$.
(c) The efficiency $\eta_{\rm int}$ of the I-OUIS model [see Eq.~(\ref{EQ:efficiencyIII_modified})] as a function of 
$\tilde \tau_{\rm m}$ for the same values of $\tilde A$ shown in (b). 
The threshold velocity is $\tilde v_0 = 0$.
 }
\label{FIG:tunable efficiency}
\end{figure*}

\begin{figure}[tbh]
\centering
\includegraphics[scale=0.23]{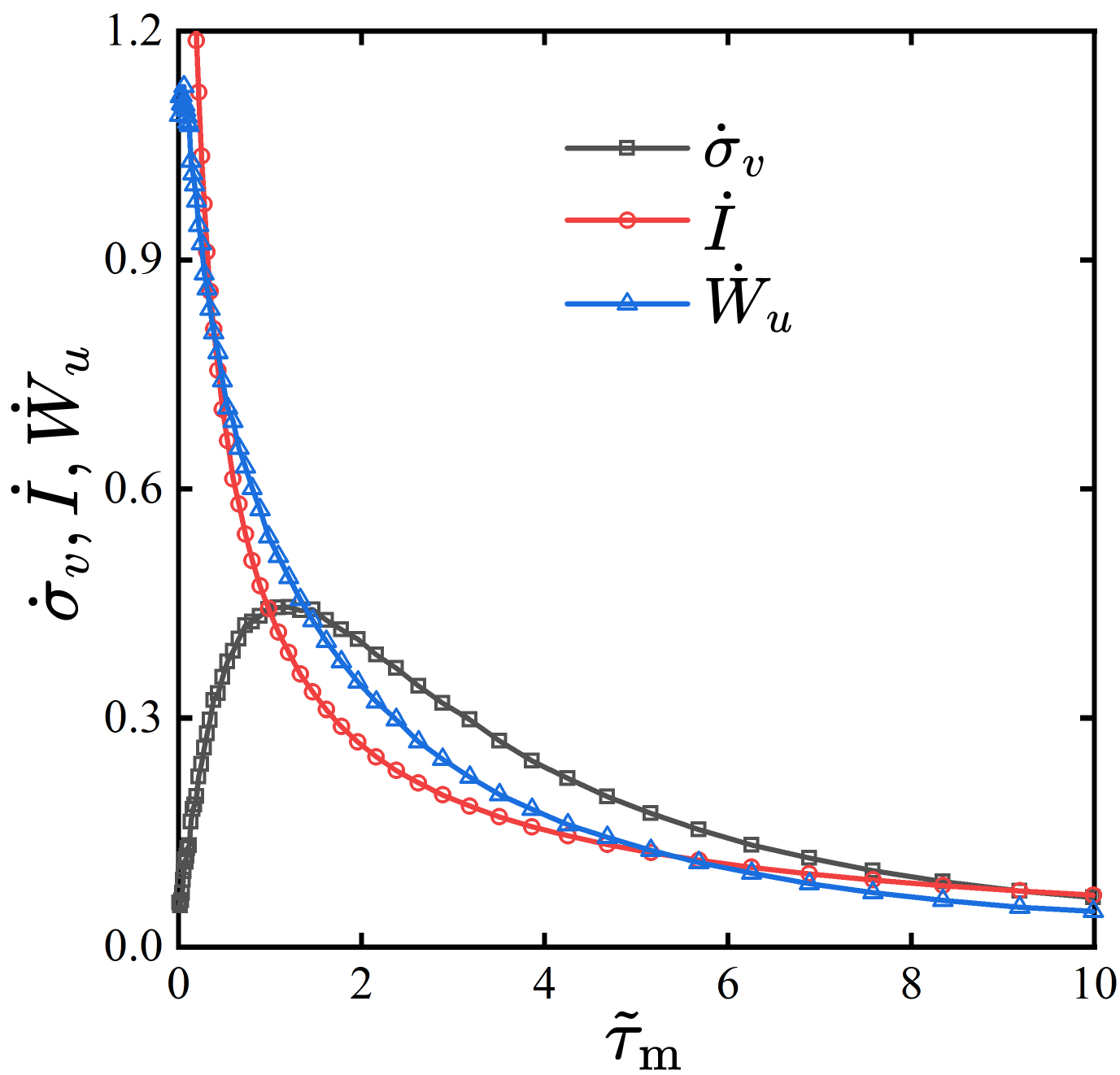}
\caption{
Contributions of dimensionless $\dot{\sigma}_v$, $\dot{I}$, and $\dot{W}_u$ to the efficiency $\eta_{\rm int}$ of the 
I-OUIS model [see Eq.~(\ref{EQ:efficiencyIII_modified})] as a function of $\tilde \tau_{\rm m}$.
The choice of parameters corresponds to that of $\tilde A =10$ (purple) in Fig.~\ref{FIG:tunable efficiency}(c).
}
\label{contributions}
\end{figure}

\section{OUIS with internal feedback control (I-OUIS)}

In the previous E-OUIS model, the friction coefficient of the AOUP was switched  
according to the measurement [see Eq.~(\ref{EQ:model2})].
Next, we consider a different type of OUIS that changes the dynamics of the driving velocity $u$ by measuring 
the center-of-mass velocity $v$, i.e., OUIS with internal feedback control (I-OUIS). 
With the same notation as before, the underdamped equation for an AOUP is given by
\begin{equation}
M\dot{v}(t)=-\Gamma [v(t)-u(t)]+ \sqrt{2 \Gamma k_{\rm B} T} \zeta (t),
\label{internalmodel}
\end{equation}
where the Gaussian white noise $\zeta (t)$ satisfies the same statistical properties as in Eq.~(\ref{EQ:white noise}).
In the I-OUIS model, the driving velocity $u$ obeys either of the following equations depending on $v$ after every time 
interval $\tau_{\rm m}$
\begin{equation}
\dot{u} (t)= \begin{dcases}
		-\frac{\beta_1 u(t)}{\tau_{\rm a}} + \frac{\beta_1 \sqrt{2A}}{\tau_{\rm a}}\xi (t), \quad v(n \tau_{\rm m}) \leq v_0, \\
		-\frac{\beta_2 u(t)}{\tau_{\rm a}} + \frac{\beta_2 \sqrt{2A}}{\tau_{\rm a}}\xi (t), \quad v(n \tau_{\rm m}) > v_0.
\end{dcases}
\end{equation}
Here, $\beta_1$ and $\beta_2$ are dimensionless coefficients and $\xi (t)$ also represents the Gaussian white noise as 
in Eq.~(\ref{EQ:white noise2}).
Different values for $\beta_1$ and $\beta_2$ (typically chosen as $\beta_1 > \beta_2$) give rise 
to the internal feedback control of the swimmer.
The dimensionless form of the I-OUIS model can be summarized as 
\begin{gather}
\frac{d \tilde v(\tilde t)}{d \tilde t} = -[\tilde v (\tilde t) - \tilde u (\tilde t)] + \tilde \zeta (\tilde t), 
\label{internalmodelnondim}
\\
\frac{d \tilde u(\tilde t)}{d \tilde t} = 
\begin{dcases}
- \frac{\beta_1 \tilde u (\tilde t)}{\tilde \tau_{\rm a}} + \frac{\beta_1 \sqrt{\tilde A}}{\tilde \tau_{\rm a}} \tilde \xi (\tilde t), & 
\quad \tilde v (n \tilde \tau_{\rm m}) \leq \tilde v_0, \\
- \frac{\beta_2 \tilde u (\tilde t)}{\tilde \tau_{\rm a}} + \frac{\beta_2 \sqrt{\tilde A}}{\tilde \tau_{\rm a}}  \tilde \xi (\tilde t), &
\quad  \tilde v (n \tilde \tau_{\rm m}) > \tilde v_0,
\end{dcases}	
\label{modelinternalactive}
\\
\langle \tilde \zeta(\tilde t) \rangle=0, \quad \langle \tilde \zeta(\tilde t) \tilde \zeta(\tilde t^\prime) \rangle
=2 \delta(\tilde t - \tilde t^\prime),
\\
\langle \tilde \xi(\tilde t) \rangle=0, \quad \langle \tilde \xi(\tilde t) \tilde \xi(\tilde t^\prime) \rangle
=2 \delta(\tilde t - \tilde t^\prime).
\label{internalmodelxi}
\end{gather}

In Figs.~\ref{FIG:non-thermal OUIS}(a) and (b), we plot the velocity distribution function 
$P(\tilde v)$ and the velocity $\langle \tilde{v} \rangle$ of the I-OUIS, respectively, 
when $(\beta_1, \beta_2, \tilde \tau_{\rm m}) = (10, 0.1, 0.01)$ for different values of $\tilde v_0$. 
Similar to the E-OUIS model, $P(\tilde v)$ becomes asymmetric with the internal 
feedback control, and $\langle \tilde{v} \rangle$ becomes finite.
Compared to Fig.~\ref{FIG:thermal_OUIS}(b), however, $\langle \tilde{v} \rangle$ is much 
larger when $\beta_2$ is made smaller, and reaches up to $\langle \tilde{v} \rangle \approx 4$ when 
$(\beta_1, \beta_2) = (10, 0.01)$ as $\tilde{v}_0$ is varied. 
In contrast to the E-OUIS case, the distribution function $P(\tilde u)$ of the driving  
velocity $u$ is no longer Gaussian and becomes highly asymmetric,
as shown in Fig.~S3(a).
Even though $P(\tilde v)$ and $P(\tilde u)$ are very different in the steady state, we have numerically 
confirmed that the corresponding average velocities coincide, 
$\langle \tilde{v} \rangle =\langle \tilde{u} \rangle$ [see Fig.~\ref{FIG:non-thermal OUIS}(b) and Fig.~S3(b)].
Although this is expected from Eqs.~(\ref{internalmodel}) and  (\ref{internalmodelnondim}), the variances of 
$P(\tilde v)$ and $P(\tilde u)$ plotted in Figs.~S3(c) and (d), respectively, are apparently different.
The results of $P(\tilde v)$ and $\langle \tilde{v} \rangle$ with a larger measurement time $\tilde \tau_{\rm m} =1$
are shown in Figs.~S4(a) and (b), respectively.

To discuss the efficiency of the I-OUIS model, we additionally need to take into account the active power 
due to the driving velocity $u$ because its average $\langle \hat{u} \rangle$ is non-zero
(unlike the E-OUIS model).
Similar to the entropy production rate $\dot{\sigma}_v$ of the E-OUIS model, the active power is  
estimated by $\dot{W}_u = \sum_i (\beta_i M/\tau_{\rm a}) \langle u \rangle_i^2 p_i/T^\ast_i$, where
$T^\ast_i =T[1+ \tilde A/(1+\tilde \tau_{\rm a}/\beta_i)]$ ($i=1,2$) [see Eq.~(\ref{effT})].
Notice that the effective friction for $u$ is given here by $\beta_i M/\tau_{\rm a}$ that is proportional 
to the mass.
Using these quantities, we consider the following modified efficiency for the I-OUIS model 
\begin{align}
\eta_{\rm int} = \frac{\dot{\sigma}_v}{\dot{I}+\dot{W}_u} 
= \frac{\sum_i \langle \tilde v -\tilde u \rangle_i \langle \tilde v \rangle_i p_i/\tilde{T}^\ast_i}
{(\tilde{I}/ \tilde \tau_{\rm m})+ \sum_i  (\beta_i/\tilde \tau_{\rm a}) \langle \tilde u \rangle_i^2 p_i/\tilde T^\ast_i},
\label{EQ:efficiencyIII_modified}
\end{align}
where $\tilde T^\ast_i = T^\ast_i/T$.

The above efficiency $\eta_{\rm int}$ is plotted as a function of  
$\tilde \tau_{\rm m}$ in Fig.~\ref{FIG:non-thermal OUIS}(c) when $\tilde v_0 = 0$. 
In the case of $(\beta_1, \beta_2) = (10, 1)$, for example, the efficiency takes a maximum value 
$\eta_{\rm int} \approx 0.23$ at around $\tilde \tau_{\rm m} \approx 1.43$. 
When $\beta_2$ is decreased to $(\beta_1, \beta_2) = (10, 0.01)$, the efficiency increases 
significantly up to $\eta_{\rm int} \approx 0.72$, and it decays slowly even for large 
$\tilde \tau_{\rm m}$ values.
In Fig.~S5, we have performed the simulation of the I-OUIS with the other choices of 
parameters for which we have fixed $(\beta_1, \tilde \tau_{\rm m}) = (1, 0.01)$ and systematically changed 
$\beta_2$.
The overall behavior is robust to the parameter choice as long as $\tau_{\rm a}$ and $\tilde A$ are fixed.

So far, the strength of active fluctuation in Eq.~(\ref{modelinternalactive}) has been fixed to 
$\tilde A=1$.
Choosing $(\beta_1, \beta_2, \tilde \tau_{\rm m}) = (10, 0.1, 0.01)$ as before, 
we plot in Fig.~\ref{FIG:tunable efficiency}(a), (b), and (c) the distribution function $P(\tilde{v})$, the velocity 
$\langle \tilde v \rangle$, and the efficiency $\eta_{\rm int}$, respectively, when $\tilde A$ is 
varied from $0.1$ to $10$.
Generally, $\langle \tilde v \rangle$ becomes larger and the range of $\tilde v_0$ that gives finite 
$\langle \tilde v \rangle$ becomes wider as $\tilde A$ is increased. 
For example, the maximum velocity exceeds $\langle \tilde v \rangle \approx 7$ and the maximum 
efficiency can be as large as $\eta_{\rm int} \approx 0.74$ when $\tilde A = 10$.
Hence, the strength of active fluctuation has a significant effect on the performance of the 
I-OUIS, and it can also control the swimming efficiency.

To discuss the efficiency $\eta_{\rm int}$ of the I-OUIS model more quantitatively, we separately plot
dimensionless $\dot{\sigma}_v$, $\dot{I}$, and $\dot{W}_u$ in Fig.~\ref{contributions} as a function of 
$\tilde{\tau}_{\rm m}$ when $\tilde A = 10$ [corresponding to the purple data in Fig.~\ref{FIG:tunable efficiency}(c)].
For small $\tilde \tau_{\rm m}$, $\dot{I}$ is larger than $\dot{W}_u$, whereas $\dot{W}_u$ becomes larger 
when $\tilde \tau_{\rm m} > 0.5$.
However, $\dot{I}$ becomes larger than $\dot{W}_u$ for large $\tilde{\tau}_{\rm m}$ values.
The contribution of $\dot{\sigma}_v$ takes a maximum value at around $\tilde \tau_{\rm m} \approx 1$. 
As a result, the efficiency is maximized with a value $\eta_{\rm int} \approx 0.74$ at around 
$\tilde \tau_{\rm m} \approx 3.6$ as shown in Fig.~\ref{FIG:tunable efficiency}(c) (the purple data).
A similar behavior can also be seen for $\tilde A = 1$ and $5$, as presented in Figs.~S6(a) and (b),
respectively. 
From these results, we find that the contribution of $\dot{W}_u$ becomes larger as $\tilde A$ is increased.

If we do not include $\dot{W}_u$ in the denominator of $\eta_{\rm int}$ in 
Eq.~(\ref{EQ:efficiencyIII_modified}) and simply estimate $\eta_{\rm ext}$ in Eq.~(\ref{efficiencyII}), 
the latter can exceed unity for certain choices of the model parameters. 
A similar situation was reported for an information engine in a non-equilibrium bath~\cite{DiLeonardo10} 
that shows an efficiency larger than unity~\cite{Saha23} (called ``pseudo-efficiency"~\cite{Datta22}). 
To be consistent with the extended second law of thermodynamics with information~\cite{Parrondo15}, 
it is necessary to consider the active power $\dot{W}_u$ resulting from the driving velocity.

\section{Discussion}

For both E-OUIS and I-OUIS models, we have assumed that the measurement and feedback control cause no energy dissipation,
and these swimmers acquire net locomotion without any external energy input.  
On the other hand, the memory spaces of these OUISs are sufficiently large so that the stored information 
entropy can increase monotonically. 
Owing to the latter assumption, the second law of thermodynamics is not violated even though these models can achieve 
a nonzero velocity under noisy conditions.
For the I-OUIS model, the second law can be stated as $\eta_{\rm int} \le 1$ or $\dot{\sigma}_v - \dot{W}_u \le \dot{I}$,
which is equivalent to that discussed in the context of information thermodynamics~\cite{Parrondo15}. 
Notice that the active power $\dot{W}_u$ vanishes in the E-OUIS model. 
We emphasize again that the maximum efficiency can be obtained by tuning the measurement time $\tau_{\rm m}$ for 
both OUIS models.

In the two OUIS models discussed in this Letter, not only the information but also the particle's inertia  
(characterized by the mass $M$) plays an essential role. 
In the presence of inertia, the unidirectional motion of the particles is maintained and serves as a memory effect.
Since we chose $\alpha_1 > \alpha_2$ and $\beta_1 > \beta_2$ for the E-OUIS and I-OUIS models, respectively, 
the inertia effect is weak when $v < v_0$ and it is strong when $v > v_0$.
Since the state with weak inertia can be partially converted to that with strong inertia owing to the switching mechanism, nonzero 
center-of-mass velocity $v$ can be maintained by positive feedback.
The bimodal asymmetric velocity distributions $P(v)$ shown in Figs.~\ref{FIG:thermal_OUIS}(a) and \ref{FIG:non-thermal OUIS}(a) 
originate from the superposition of a Gaussian distribution reflecting the weak inertia case ($v < v_0$) 
and a biased distribution reflecting the strong inertia case ($v > v_0$).
A similar bimodal velocity distribution was also reported in Ref.~\cite{Huang20}.

Corresponding to the I-OUIS, we give some numbers for the model parameters.
Recently, the run-and-tumble motion of \textit{E.\ coli} has been quantitatively measured by using intermediate 
scattering functions and the renewal theory~\cite{Kurzthaler24,Zhao24}. 
For a wild-type \textit{E.\ coli}, the running and tumbling times were estimated to be $\tau_{\rm R}=2.39$\,s and 
$\tau_{\rm T}=0.38$\,s, respectively, and the average self-propulsion velocity was $u_0=15.95$\,$\mu$m/s.  
If we identify $\tau_{\rm R}$ as the persistence time $\tau_{\rm a}$ in our OUIS models, the strength of active fluctuation  
can be estimated as $A \approx 4 \times 10^{-10}$\,m$^2$/s.
Since the ratio $\beta_1/\beta_2$ can also be estimated from the ratio $\tau_{\rm R}/\tau_{\rm T} \approx 6$, the choice 
$(\beta_1, \beta_2)=(10,1)$ in Fig.~\ref{FIG:non-thermal OUIS} is realistic for a wild-type \textit{E.\ coli}.
Experimentally, the running and tumbling times can be controlled by adding a chemical inducer~\cite{Kurzthaler24,Zhao24}.

In this work, we have implemented a process of measurement and feedback control in an AOUP.
However, the concept of information swimmer is more general. 
For example, Kumar \textit{et al.} proposed a Brownian inchworm model of a self-propelled elastic 
dimer~\cite{Kumar08,Baule08}.
In their model, the crucial mechanism is the position-dependent friction coefficients of the two particles.  
Although such a mechanism was not regarded as an explicit feedback control operation, the proposed Brownian 
inchworm model offers a typical example of informational active matter~\cite{Li25}.  
Currently, we are constructing a general framework of information swimmers using the  
statistical formulation of Onsager-Machlup variational principle~\cite{Yasuda24}.

\section{Summary}

In this Letter, we have performed numerical simulations of the information swimmers based on the active Ornstein-Uhlenbeck 
model; E-OUIS with external feedback control and I-OUIS with internal feedback control.
Both OUIS models exhibit nonzero average velocities in the steady state, and their statistical properties as well as the efficiencies 
have been discussed.
For both models, maximum efficiency can be obtained by tuning the measurement time, and hence the information entropy 
plays an essential role in their locomotion.
Depending on the choice of the model parameters, I-OUIS can achieve a larger average velocity and higher efficiency 
than those of E-OUIS when the active fluctuation is large. 
Our models provide an important step toward understanding informational transport in biological systems.

\acknowledgments
We thank Y.\ Hosaka for the useful discussion.
Z.H.\ and S.K.\ acknowledge the support by the National Natural Science Foundation of China 
(Nos.\ 12104453, 12274098, and 12250710127).
S.K.\ acknowledges the startup grant of Wenzhou Institute, 
University of Chinese Academy of Sciences (No.\ WIUCASQD2021041). 
K.Y\ and S.K.\ acknowledge the support by the Japan Society for the Promotion of Science (JSPS) Core-to-Core 
Program ``Advanced core-to-core network for the physics of self-organizing active matter" (No.\ JPJSCCA20230002).
Z.H.\ and Z.Z.\ contributed equally to this work.

\end{document}


\title{\textcolor{black}{Supplementary Information} for ``Ornstein-Uhlenbeck information swimmers with external and internal feedback controls''}

\author{Zhanglin Hou}
\affiliation{Wenzhou Institute, University of Chinese Academy of Sciences, 
Wenzhou, Zhejiang 325001, China} 
\affiliation{Oujiang Laboratory, Wenzhou, Zhejiang 325000, China}

\author{Ziluo Zhang}
\affiliation{Wenzhou Institute, University of Chinese Academy of Sciences, 
Wenzhou, Zhejiang 325001, China}
\affiliation{Institute of Theoretical Physics, Chinese Academy of Sciences, 
Beijing 100190, China}

\author{Jun Li}
\affiliation{Department of Physics, Wenzhou University, Wenzhou, Zhejiang 325035, China}
\affiliation{Wenzhou Institute, University of Chinese Academy of Sciences, 
Wenzhou, Zhejiang 325001, China}

\author{Kento Yasuda}
\affiliation{Research Institute for Mathematical Sciences, 
Kyoto University, Kyoto 606-8502, Japan}

\author{Shigeyuki Komura}\email{Corresponding author: komura@wiucas.ac.cn}
\affiliation{Wenzhou Institute, University of Chinese Academy of Sciences, 
Wenzhou, Zhejiang 325001, China} 
\affiliation{Oujiang Laboratory, Wenzhou, Zhejiang 325000, China}
\affiliation{Department of Chemistry, Graduate School of Science,
Tokyo Metropolitan University, Tokyo 192-0397, Japan}

\maketitle


\newcommand{\ttilde}{\tilde{t}}
\newcommand{\utilde}{\tilde{u}}
\newcommand{\vtilde}{\tilde{v}}
\newcommand{\Atilde}{\tilde{A}}
\newcommand{\phitilde}{\tilde{\phi}}
\newcommand{\psitilde}{\tilde{\psi}}
\newcommand{\tautilde}{\tilde{\tau}}
\newcommand{\plaind}{\mathrm{d}}
\newcommand{\dbar}{\plaind\mkern-6mu\mathchar'26}
\newcommand{\imag}{\mathring{\imath}}
\renewcommand{\exp}[1]{\mathchoice{\mathrm{e}^{#1}}{\operatorname{exp}\left(#1\right)}{\operatorname{exp}\left(#1\right)}{\operatorname{exp}\left(#1\right)}}
\newcommand{\ave}[1]{\left\langle #1 \right\rangle}
\newcommand\ptwiddle[1]{\mathord{\mathop{#1}\limits^{\scriptscriptstyle(\sim)}}}
\newcommand{\Hermite}[1]{H\!e_{#1}}
\newcommand{\eigenfunc}{f}
\newcommand{\dint}[1]{\mathchoice{\!\plaind#1\,}{\!\plaind#1\,}{\!\plaind#1\,}{\!\plaind#1\,}}

\section{Effective temperature of an active Ornstein-Uhlenbeck particle}
\label{App:effective_temperature}

The effective temperature of an active Ornstein-Uhlenbeck particle (AOUP) is obtained by using its 
equal-time velocity correlation function $\langle \tilde{v}^2\rangle$. 
Here, we briefly present our mathematical scheme to obtain it, while the details of a similar approach can be found 
in Refs.~\cite{Garcia-MillanPruessner:2021,ZYK24}.
The dimensionless equation of motion for an AOUP can be written as
\begin{gather}
\frac{d\tilde v(\tilde t)}{d\tilde t} = 
- [\tilde v(\tilde t) - \tilde u(\tilde t)]+\tilde \zeta(\tilde t), 
\label{externalmodel}
\\
\frac{d \tilde u(\tilde t)}{d \tilde t}=-\frac{\tilde u(\tilde t)}{\tilde \tau_{\rm a}} + \frac{\sqrt{\tilde A}}{\tilde \tau_{\rm a}} \tilde \xi(\tilde t),
\label{EQ:colourednondimension}
\\
\langle \tilde \zeta(\tilde t) \rangle=0, \quad \langle \tilde \zeta(\tilde t) \tilde \zeta(\tilde t^\prime) \rangle
=2 \delta(\tilde t - \tilde t^\prime),
\\
\langle \tilde \xi(\tilde t) \rangle=0, \quad \langle \tilde \xi(\tilde t) \tilde \xi(\tilde t^\prime) \rangle
=2 \delta(\tilde t - \tilde t^\prime).
\label{externalmodelnoisexi}
\end{gather}
The above equations can be converted to the following Fokker-Planck equation for the probability distribution function
$P(\tilde{v},\tilde{u};t)$:
\begin{equation}
\frac{\partial P}{\partial t}=
\frac{\partial^2 P}{\partial \tilde{v}^2}
+\frac{\partial}{\partial \tilde{v}}[(\vtilde-\utilde)P]
+\frac{\tilde{A}}{\tilde{\tau}_a}
\frac{\partial^2 P}{\partial \tilde{u}^2}
+\frac{\partial}{\partial \tilde{u}}(\utilde P).
\end{equation}

We use the Doi-Peliti field theory to obtain the correlation function. 
Using the definition $L=\sqrt{\Atilde/\tautilde_a}$, we first introduce the fields as 
\begin{align}
\ptwiddle{\phi}(\vtilde,t)&=\int \frac{d\omega}{2\pi} \, e^{-{\rm i} \omega t}\sum_{n=0}^\infty\ptwiddle{f}_n(\vtilde),  
\\
\ptwiddle{\psi}(\utilde,t)&=\int \frac{d\omega}{2\pi} \, e^{-{\rm i} \omega t} \frac{1}{L}\sum_{n=0}^\infty\ptwiddle{f}_n(\utilde/L),
\end{align}
where $\phi$ and $\psi$ are the annihilation fields, whereas $\phitilde$ and $\psitilde$ are the Doi-shifted 
creation fields~\cite{Garcia-MillanPruessner:2021,ZYK24}.
The functions $\ptwiddle{f}_n$ are defined as 
\begin{align}
\eigenfunc_n(x)&=e^{-x^2/2}\Hermite{n}(x),
\\
\tilde{\eigenfunc}_n(x)&=\frac{1}{\sqrt{2\pi}n!}\Hermite{n}(x),
\end{align}
where $\Hermite{n}(x)$  is the $n$-th order of the probabilist's 
Hermite polynomial~\cite{Garcia-MillanPruessner:2021,ZYK24}.
Their orthogonality relation is given by 
\begin{equation}
 \int dx \, \eigenfunc_n(x/L) \tilde{\eigenfunc}_m(x/L)=L\delta_{nm},
\end{equation}
where $\delta_{nm}$ is the Kronecker delta and $\eigenfunc_n(x/L)$ is the eigenfunction of the following operator:
\begin{equation}
L^2\partial_x^2 f_n(x/L)+\partial_x [x f_n(x/L)]=-nf_n(x/L).
\end{equation}

The two-point correlation function under the given initial state $(\vtilde_0,\utilde_0,t_0)$ is written as 
\begin{align}
P(\vtilde,\utilde,t|\vtilde_0,\utilde_0,t_0)&=\big \langle \phi(\vtilde,t)\psi(\utilde,t)\phitilde(\vtilde_0,t_0)\psitilde(\utilde_0,t_0) \big \rangle \\
    &=\frac{1}{L^2}\sum_{n,n',m,m'}\eigenfunc_n(\vtilde)\eigenfunc_m(\utilde/L)
    \big \langle \phi_n(t)\psi_m(t)\phitilde_{n'}(t_0)\psitilde_{m'}(t_0) \big \rangle
    \tilde{\eigenfunc}_{n'}(\vtilde)\tilde{\eigenfunc}_{m'}(\utilde/L).
\label{eq:corr_func}
\end{align}
Then, the equal-time correlation function is calculated by 
\begin{equation}
\label{eq:vtilde_corr}
\langle\tilde{v}^2\rangle=\lim_{t_0 \rightarrow -\infty} \int d\tilde{v} \int d\tilde{u} \, \vtilde^2 P(\vtilde,\utilde,t|\vtilde_0,\utilde_0,t_0).
\end{equation}
We substitute Eq.~(\ref{eq:corr_func}) into Eq.~(\ref{eq:vtilde_corr}) and use the relation $x^2=\Hermite{2}(x)+\Hermite{0}(x)$.
Applying the orthogonality conditions of the Hermite polynomials, we finally obtain the equal-time correlation as 
\begin{equation}
\langle\tilde{v}^2\rangle=1+\frac{\tilde{A}}{1+\tilde{\tau}_a}.
\end{equation}
Here, the second term corresponds to the temperature increase due to the Ornstein-Uhlenbeck noise.

\section{Additional numerical data}

\begin{figure}[tbh]
\centering
\includegraphics[scale=0.25]{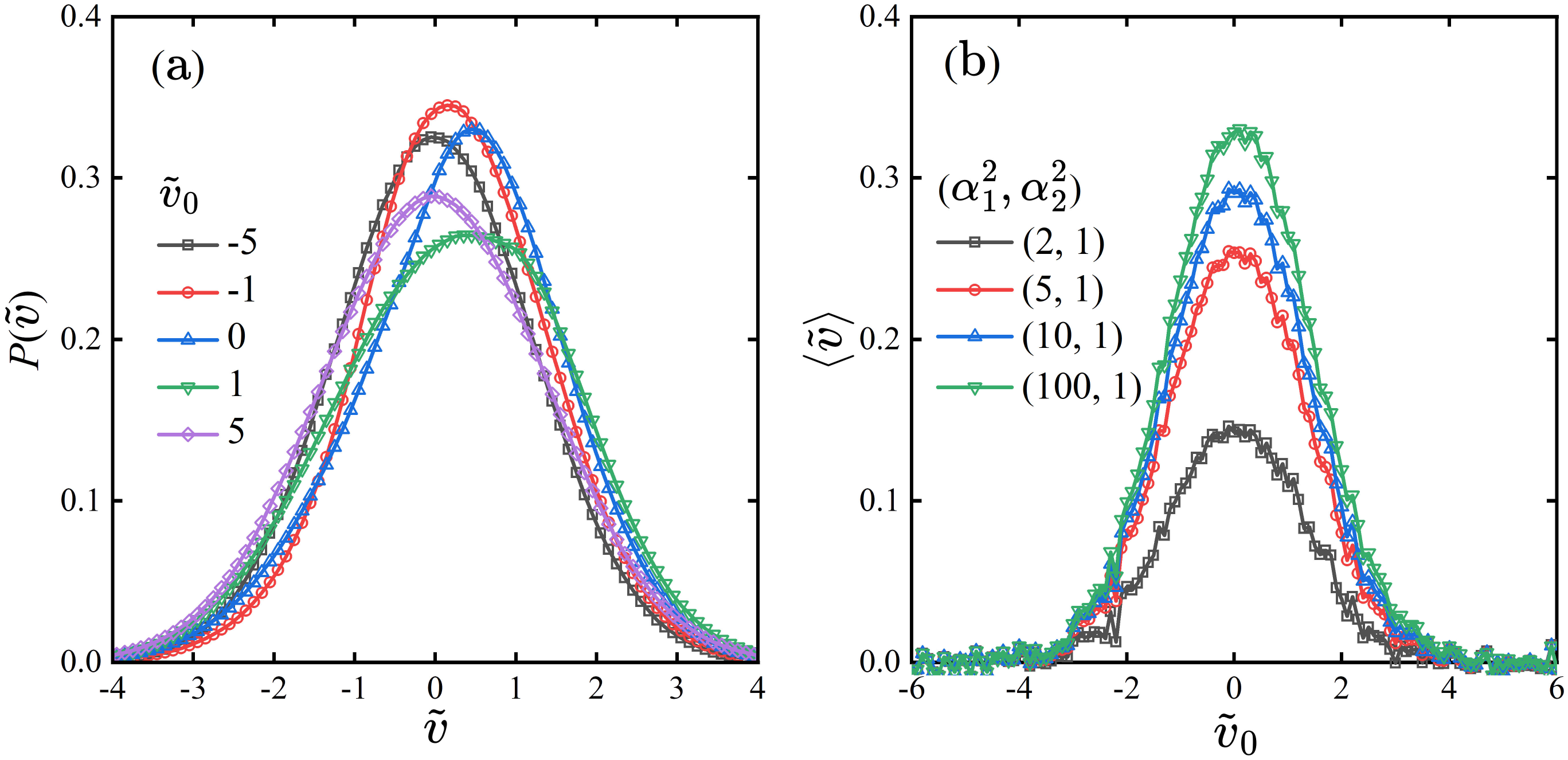}
\caption{
Various statistical properties of the E-OUIS model with external feedback control [see Eqs.~(5)--(8)].
The fixed parameters are $\tilde \tau_{\rm a} = 1$, $\tilde A = 1$, and $\tilde \tau_{\rm m} = 1$.
(a) The steady-state velocity distribution function $P(\tilde v)$ when the threshold velocity is changed within the range of $-5 \le \tilde{v}_0 \le 5$.
Here, we choose $(\alpha^2_1,\alpha^2_2)=(10,1)$.
(b) The steady-state average velocity $\langle \tilde v \rangle$ as a function of the threshold velocity $\tilde v_0$ when 
$(\alpha^2_1,\alpha^2_2)=(2,1)$, $(5,1)$, $(10,1)$ and $(100,1)$.
}
\label{FIG:S1}
\end{figure}

\begin{figure}[tbh]
\centering
\includegraphics[scale=0.25]{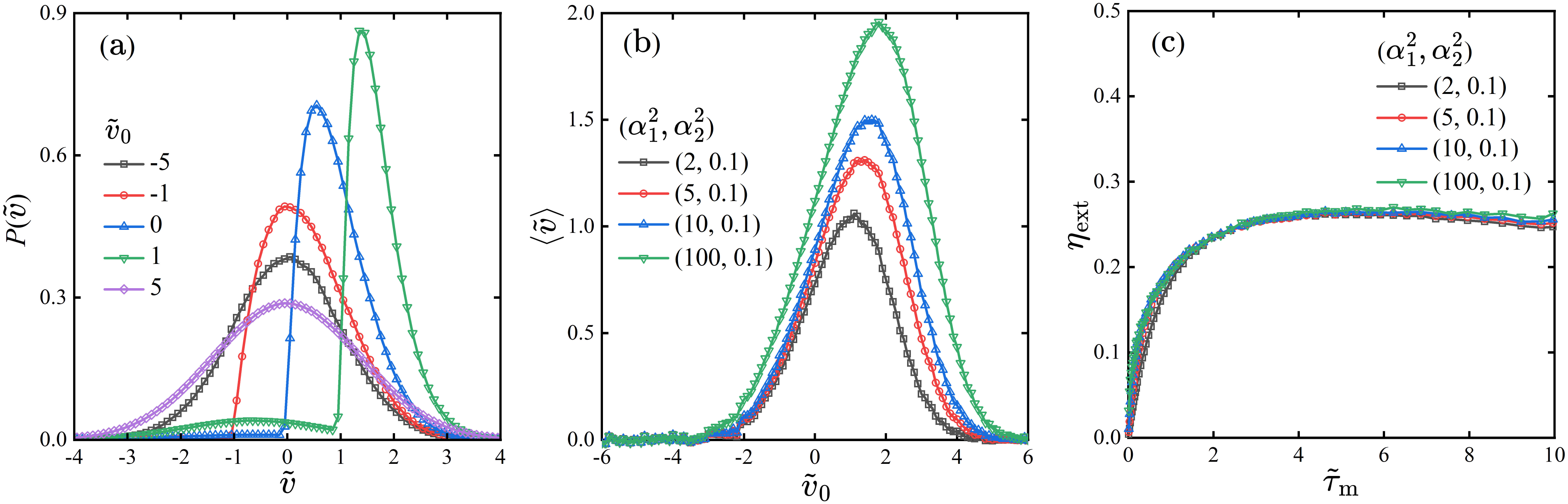}
\caption{
Various statistical properties of the E-OUIS model with external feedback control [see Eqs.~(5)--(8)].
The fixed parameters are $\tilde \tau_{\rm a} = 1$ and $\tilde A = 1$.
(a) The steady-state velocity distribution function $P(\tilde v)$ when the threshold velocity is changed within the range of $-5 \le \tilde{v}_0 \le 5$.
Here, we choose $(\alpha^2_1,\alpha^2_2)=(10,0.1)$ and the measurement time interval is $\tilde \tau_{\rm m} = 0.01$.
(b) The steady-state average velocity $\langle \tilde v \rangle$ as a function of the threshold velocity $\tilde v_0$ when 
$(\alpha^2_1,\alpha^2_2)=(2,0.1)$, $(5,0.1)$, $(10,0.1)$, $(100,0.1)$ and $\tilde \tau_{\rm m} = 0.01$.
(c) The efficiency $\eta_{\rm ext}$ of the E-OUIS model with external feedback control [see Eq.~(10)] as a function of 
$\tilde \tau_{\rm m}$ for the same combinations of $(\alpha^2_1,\alpha^2_2)$ shown in (b). 
The threshold velocity is $\tilde v_0 = 0$.
}
\label{FIG:S2}
\end{figure}

\begin{figure}[tbh]
\centering
\includegraphics[scale=0.25]{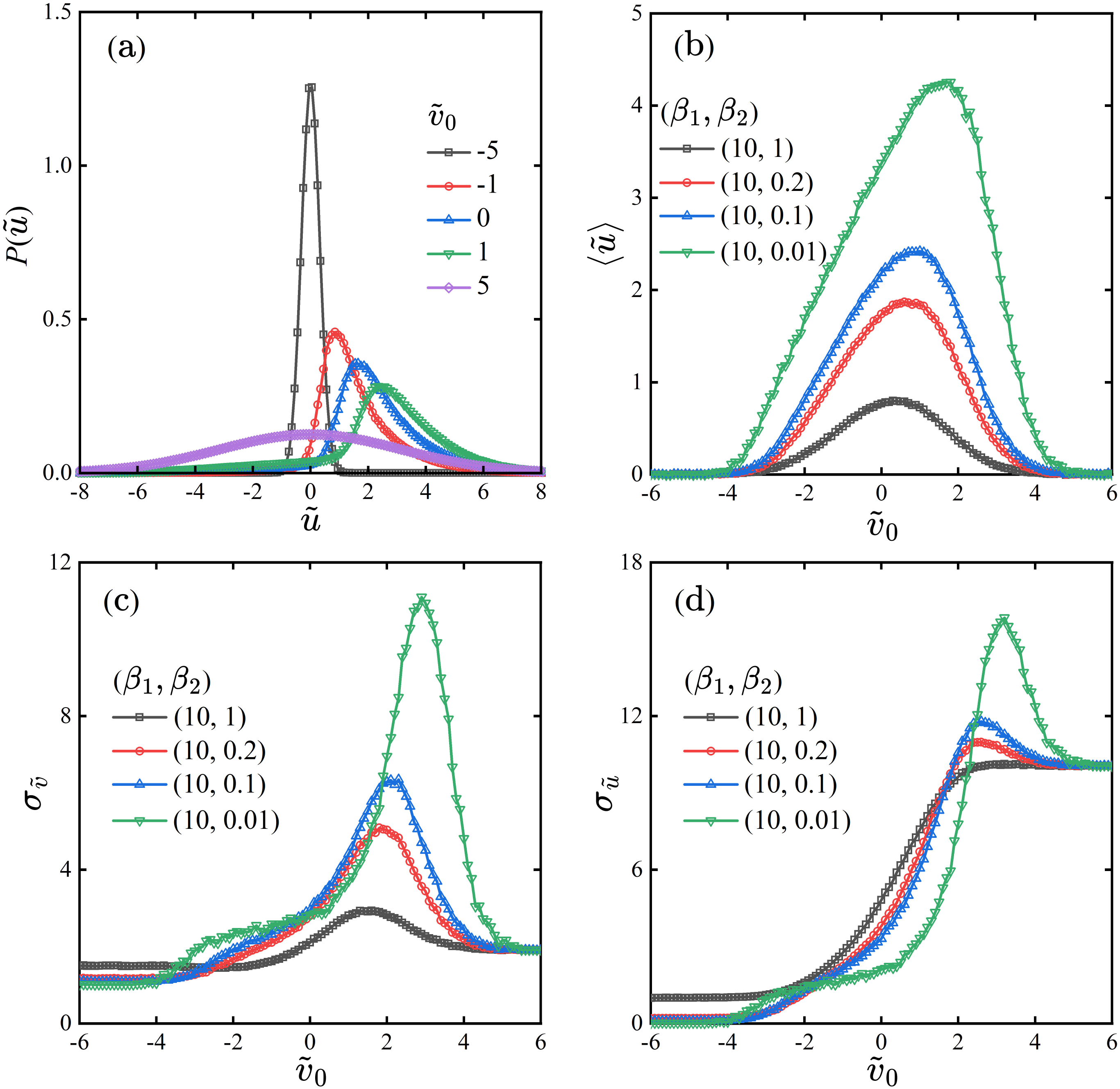}
\caption{
Various statistical properties of the I-OUIS model with internal feedback control [see Eqs.~(13)--(16)].
The fixed parameters are $\tilde \tau_{\rm a} = 1$, $\tilde A = 1$, and $\tilde \tau_{\rm m} = 0.01$.
(a) The steady-state self-propulsion velocity distribution function $P(\tilde u)$ when the threshold velocity is changed within the range of $-5 \le \tilde{v}_0 \le 5$.
Here, we choose $(\beta_1,\beta_2)=(10,0.1)$.
(b) The steady-state average self-propulsion velocity $\langle \tilde u \rangle$ as a function of the threshold velocity $\tilde v_0$ when 
$(\beta_1,\beta_2)=(10,1)$, $(10,0.2)$, $(10,0.1)$ and $(10,0.01)$.
(c) The variance $\sigma_{\tilde{v}}$ of the steady-state velocity distribution function $P(\tilde v)$ in Fig.~2 as a function of $\tilde v_0$.
(d) The variance $\sigma_{\tilde{u}}$ of the steady-state self-propulsion velocity distribution function $P(\tilde u)$ in Fig.~S3 as a function of $\tilde v_0$.
}
\label{FIG:S3}
\end{figure}

\begin{figure}[tbh]
\centering
\includegraphics[scale=0.25]{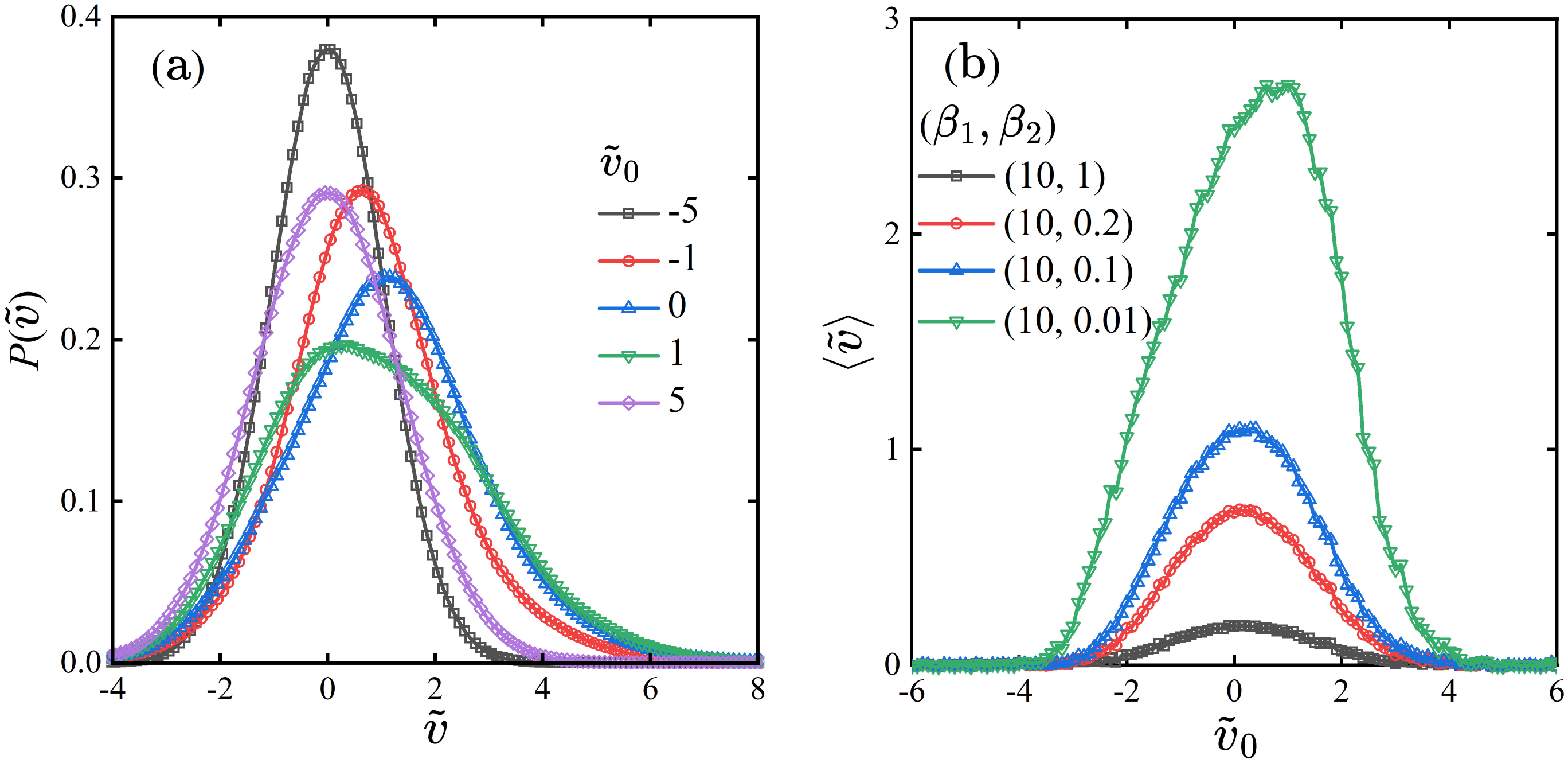}
\caption{
Various statistical properties of the I-OUIS model with internal feedback control [see Eqs.~(13)--(16)].
The fixed parameters are $\tilde \tau_{\rm a} = 1$, $\tilde A = 1$, and $\tilde \tau_{\rm m} = 1$.
(a) The steady-state velocity distribution function $P(\tilde v)$ when the threshold velocity is changed within the range of $-5 \le \tilde{v}_0 \le 5$.
Here, we choose $(\beta_1,\beta_2)=(10,0.1)$. 
(b) The steady-state average velocity $\langle \tilde v \rangle$ as a function of the threshold velocity $\tilde v_0$ when 
$(\beta_1,\beta_2)=(10,1)$, $(10,0.2)$, $(10,0.1)$ and $(10,0.01)$.
}
\label{FIG:S4}
\end{figure}

\begin{figure}[tbh]
\centering
\includegraphics[scale=0.25]{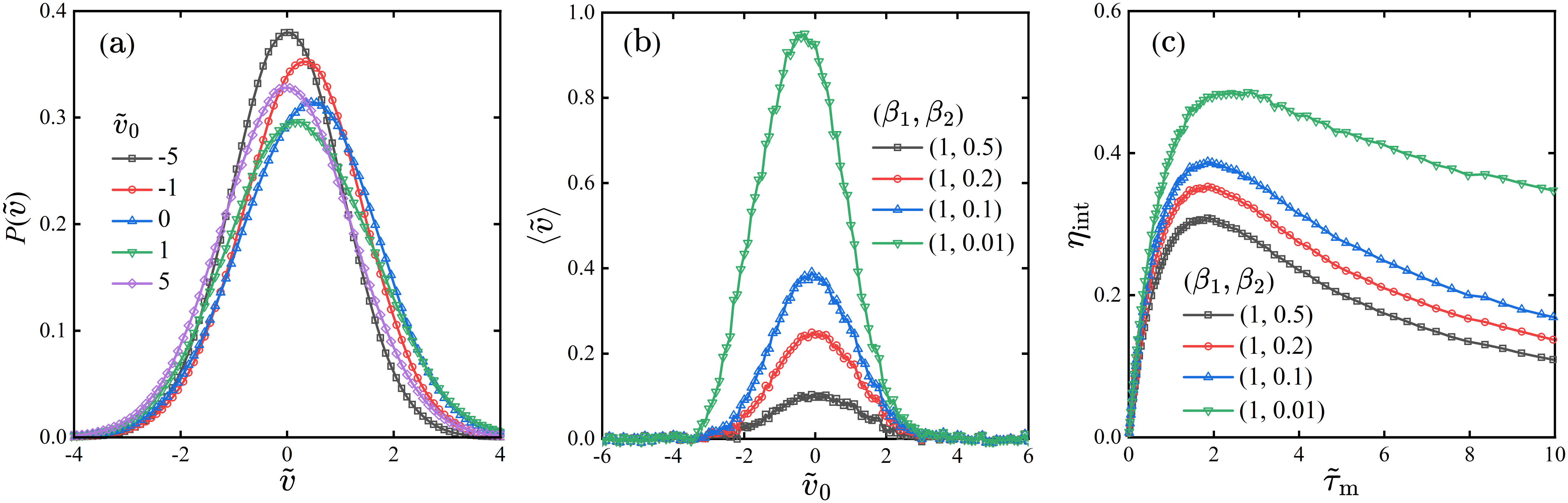}
\caption{
Various statistical properties of the I-OUIS model with internal feedback control [see Eqs.~(13)--(16)].
The fixed parameters are $\tilde \tau_{\rm a} = 1$ and $\tilde A = 1$.
(a) The steady-state velocity distribution function $P(\tilde v)$ when the threshold velocity is changed within the range of $-5 \le \tilde{v}_0 \le 5$.
Here, we choose $(\beta_1,\beta_2)=(1,0.1)$ and the measurement time interval is $\tilde \tau_{\rm m} = 0.01$.
(b) The steady-state average velocity $\langle \tilde v \rangle$ as a function of the threshold velocity $\tilde v_0$ when 
$(\beta_1,\beta_2)=(1,0.5)$, $(1,0.2)$, $(1,0.1)$, $(1,0.01)$ and $\tilde \tau_{\rm m} = 0.01$.
(c) The efficiency $\eta_{\rm int}$ of the I-OUIS model with internal feedback control [see Eq.~(17)] as a function of 
$\tilde \tau_{\rm m}$ for the same combinations of $(\beta_1,\beta_2)$ shown in (b). 
The threshold velocity is $\tilde v_0 = 0$.
}
\label{FIG:S5}
\end{figure}

\begin{figure}[tbh]
\centering
\includegraphics[scale=0.25]{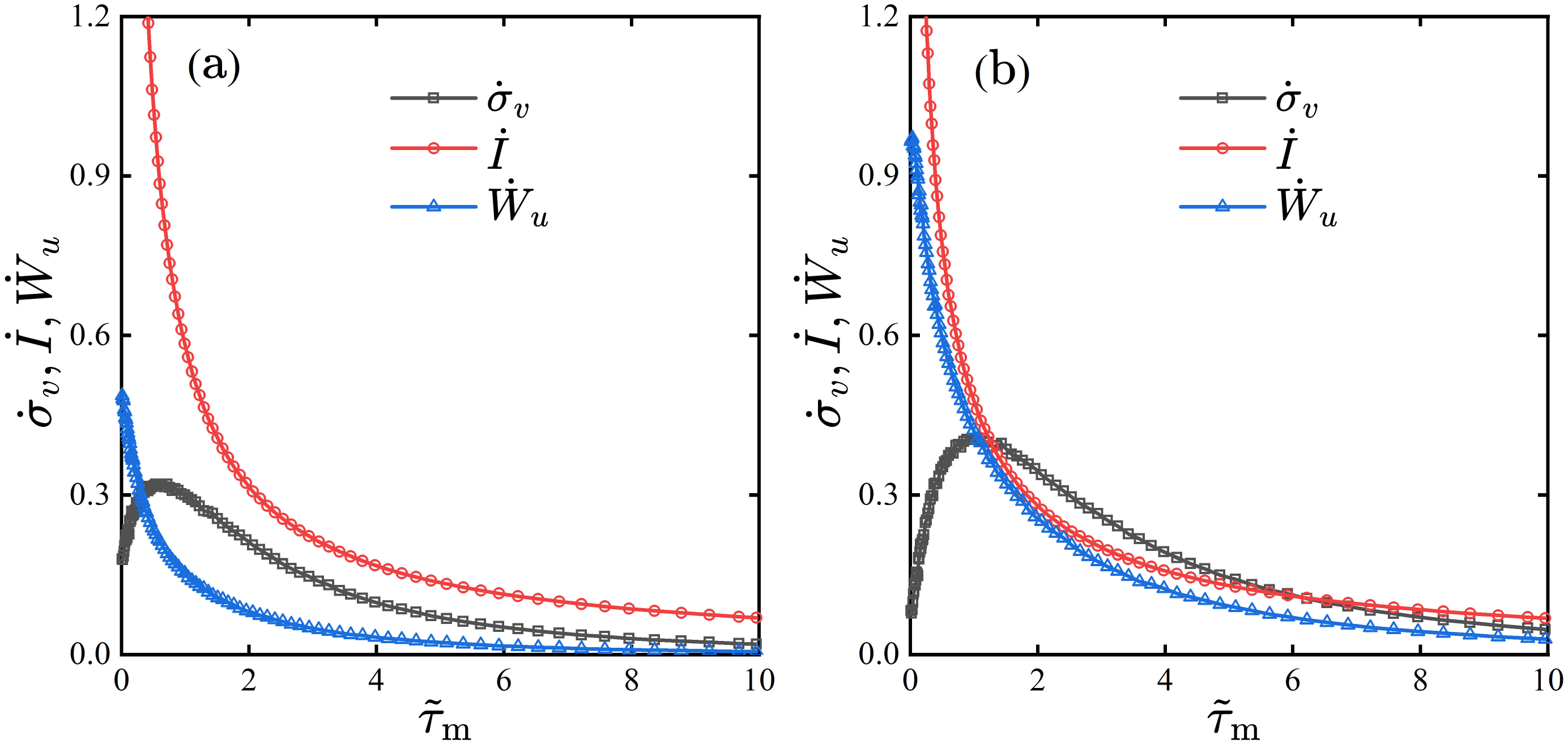}
\caption{
Contributions of dimensionless $\dot{\sigma}_v$, $\dot{I}$, and $\dot{W}_u$ to the efficiency $\eta_{\rm int}$ of the I-OUIS model
[see Eq.~(17)] as a function of $\tilde \tau_{\rm m}$.
The choice of parameters corresponds to that of (a) $\tilde A =1$ (red) and (b) $\tilde A =5$ (green) in Fig.~3(c).
}
\label{FIG:S7}
\end{figure}

\clearpage